\documentclass[conference]{IEEEtran}
\IEEEoverridecommandlockouts
\usepackage{cite}
\usepackage{amsmath,amssymb,amsfonts}
\usepackage{algorithmic}
\usepackage{graphicx}
\usepackage{textcomp}
\usepackage{xcolor}
\usepackage{subfigure}
\usepackage[outdir=./]{epstopdf}
\usepackage{float}
\usepackage{pgf,tikz}
\def\BibTeX{{\rm B\kern-.05em{\sc i\kern-.025em b}\kern-.08em
    T\kern-.1667em\lower.7ex\hbox{E}\kern-.125emX}}
\begin{document}

\title{Generalized LDPC Codes \\ with Convolutional Code Constraints
}

\author{
\IEEEauthorblockN{Muhammad Umar Farooq$^\dag$, Saeedeh Moloudi$^\ddag$, and Michael Lentmaier$^\dag$}
\IEEEauthorblockA{$\dag$Department of Electrical and Information
  Technology, Lund University, Lund, Sweden \\
  $\ddag$Ericsson Research, Mobilv\"agen 12, 223 62 Lund, Sweden \\              \{muhammad.umar\_farooq,michael.lentmaier\}@eit.lth.se, saeedeh.moloudi@ericsson.com}\\
              \thanks{This work was supported in part by the Swedish Research Council (VR) under grant \#2017-04370.}\vspace*{-1cm}
}



\maketitle

\begin{abstract}
Braided convolutional codes (BCCs) are a class of spatially coupled turbo-like codes that can be described by a $(2,3)$-regular compact graph. In this paper, we introduce a family of $(d_v,d_c)$-regular GLDPC codes with convolutional code constraints (CC-GLDPC codes), which form an extension of classical BCCs to arbitrary regular graphs. In order to characterize the performance in the waterfall and error floor regions, we perform an analysis of the density evolution thresholds as well as the finite-length ensemble weight enumerators and minimum distances of the ensembles. In particular, we consider various ensembles of overall rate $R=1/3$ and $R=1/2$ and study the trade-off between variable node degree and strength of the component codes. We also compare the results to corresponding classical LDPC codes with equal degrees and rates.  It is  observed that for the considered LDPC codes with variable node degree $d_v>2$, we can find a CC-GLDPC code with smaller $d_v$ that offers similar or better performance in terms of BP and MAP thresholds at the expense of a negligible loss in the minimum distance. 
\end{abstract}


\section{Introduction}
Turbo codes and low-density parity-check (LDPC) codes are widely used forward error correction techniques in many communication applications. For LDPC convolutional codes \cite{JimenezLDPCCC,LentmaierTransITOct2010}, also known as spatially coupled LDPC (SC-LDPC) codes, it has been proved that the threshold of efficient belief propagation (BP) decoding saturates to the threshold of an optimal maximum a-posteriori probability (MAP) decoder \cite{Kudekar_ThresholdSaturation,Yedla2014}. Spatially coupled turbo-like codes were introduced in \cite{Moloudi_SCTC_Journal}, where it was proved that threshold saturation also occurs for this class of codes. It was observed that turbo-like codes with good BP thresholds tend to have weaker MAP thresholds and minimum distance \cite{Moloudi_SCTC_Journal,MOSCTCNewtrade}. Braided convolutional codes (BCCs)\cite{Costello2016}, which are characterized by $(2,3)$-regular graphs, have better MAP thresholds and distances than parallel concatenated convolutional codes that suffer from degree-one variable nodes. In combination with spatial coupling, ensembles with good MAP thresholds and low error floors are able to simultaneously approach capacity and achieve very low error floor thanks to the threshold saturation phenomenon \cite{MOSCTCNewtrade}.

In principle, it is possible to improve the threshold and minimum distance of an SC-LDPC ensemble by increasing the variable node degree. For finite block lengths, however, ensembles with stronger component codes can have advantages \cite{Liu2018a}, since larger variable node degrees increase the number of short cycles in the factor graph, which negatively impacts the performance of a BP decoder. On the other hand, it has been observed in \cite{MOSCTCNewtrade} that, due to the stronger component codes at the constraint nodes, spatially coupled turbo-like ensembles can achieve excellent decoding thresholds and minimum distances with low variable node degrees.

In this work, our aim is to gain a better understanding of the general trade-off between increasing the variable node or the strength of the component codes. For this purpose, we introduce a family of $(d_v,d_c)$-regular generalized LDPC codes with convolutional code constraints (CC-GLDPC codes), which form an extension of classical BCCs to arbitrary regular graphs and allow for a one-to-one comparison with the corresponding $(d_v,d_c)$-regular LDPC code ensembles. As examples we consider $(2,3)$, $(4,6)$ and $(6,9)$ graphs of rate $R=1/3$ as well as $(2,4)$, $(3,6)$ and $(4,8)$ graphs of rate $R=1/2$, based on component code trellises with $2$, $4$ and $8$ states. For these ensembles we determine the BP thresholds (with and without spatial coupling), MAP thresholds and minimum distances and compare them with the corresponding LDPC code ensembles.

\section{Code Ensembles} \label{sec:CE}
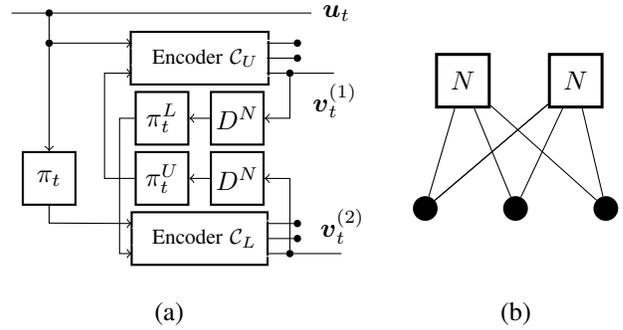
\begin{figure}[t]
	\centering
	\begin{tikzpicture}
        [var/.style={circle,draw=black!100,fill=black!100,thick,
        inner sep=0pt,minimum size=.7mm},
        check/.style={rectangle,draw=black!100,fill=black!0,thick,
        inner sep=0pt,minimum size=7mm},
        enc/.style={rectangle,draw=black!100,fill=black!0,thick,
        inner sep=0pt,minimum size=7mm},
        invisible/.style={circle,draw=black!0,fill=black!0,thick,
        inner sep=0pt,minimum size=.001mm},
        var1/.style={circle,draw=black!100,fill=black!100,thick,
        inner sep=0pt,minimum size=3mm},
        check1/.style={rectangle,draw=black!100,fill=black!0,line width=.4mm,
        inner sep=0pt,minimum size=7mm}]
        [var]
        \def\recpath{-- +(1.8,0) -- +(1.8,.7) -- +(0,.7) -- cycle}
        
        
        
        \node (vu1_in) at (-.89,2.6)  [invisible,label=left:]{};
        \node (vu2_in) at (-.89,2.2)  [invisible,label=left:]{};
        
        \node (vl1_in) at (-.89,.2)  [invisible,label=left:]{};
        \node (vl2_in) at (-.89,-.2)  [invisible,label=left:]{};
        
        \node (vu1_out) at (.88,2.6)  [invisible,label=left:]{};
        \node (vu2_out) at (.88,2.4)  [invisible,label=left:]{};
        \node (vu3_out) at (.88,2.2)  [invisible,label=left:]{};
        
        \node (vl1_out) at (.88,.2)  [invisible,label=left:]{};
        \node (vl2_out) at (.88,0)  [invisible,label=left:]{};
        \node (vl3_out) at (.88,-.2)  [invisible,label=left:]{};
        
        
        \node (piu_in) at (-.85,.8)  [invisible,label=left:]{};
        \node (piu_out) at (-.15,.8)  [invisible,label=left:]{};
        
        \node (du_in) at (.15,.8)  [invisible,label=left:]{};
        \node (du_out) at (.85,.8)  [invisible,label=left:]{};
        
        \node (pil_in) at (-.85,1.6)  [invisible,label=left:]{};
        \node (pil_out) at (-.15,1.6)  [invisible,label=left:]{};
        
        \node (dl_in) at (.15,1.6)  [invisible,label=left:]{};
        \node (dl_out) at (.85,1.6)  [invisible,label=left:]{};        
        \node (pi_in) at (-2,1.15)  [invisible,label=left:]{};
        \node (pi_out) at (-2,.45)  [invisible,label=left:]{};

        \node (vO1) at (1.5,3)  [invisible,label=right:$\boldsymbol{u}_t$]{};
        \node (vO2) at (1.8,2.2)  [invisible,label=below:$\boldsymbol{v}_t^{(1)}$]{};
        \node (vO3) at (1.9,-.2)  [invisible,label=above:$\boldsymbol{v}_t^{(2)}$]{};
        
        \node (bl1) at (-2,.2) [invisible,label=left:]{};
        \node (bpiL) at (-1.4,1.6) [invisible,label=left:]{};
        \node (bpiU) at (-1.25,.8) [invisible,label=left:]{};
        \node (bl2) at (-1.4,-.2) [invisible,label=left:]{};
        \node (bu2) at (-1.25,2.2) [invisible,label=left:]{};
        \node (bpiLo) at (1.2,1.6) [invisible,label=left:]{};
        \node (bpiUo) at (1.2,.8) [invisible,label=left:]{};

        \node (encCL) at (0,0) [check,text=white,label=left:] {IEncoder $\mathcal{C}_L $};
        \node[font=\footnotesize] [draw=black!0,text width=50,] at (.25,0) {Encoder $\mathcal{C}_L$};
        \node (encCU) at (0,2.4) [check,text=white,label=left:] {IEncoder $\mathcal{C}_U $};
        \node[font=\footnotesize] [draw=black!0,text width=50,] at (.25,2.4) {Encoder $\mathcal{C}_U$};
        \node (permU) at (-.5,.8) [check,label=left:] { $\pi_t^U$};
        \node (permL) at (-.5,1.6) [check,label=left:] { $\pi_t^L$};
        \node (delU) at (0.5,.8) [check,label=left:] { $D^N$};
        \node (permL) at (0.5,1.6) [check,label=left:] { $D^N$};
        \node (perm) at (-2,.8) [check,label=left:] { $\pi_t$};
        
        \node (vl1) at (1.3,.2)  [var,label=left:]{};
        \node (vl2) at (1.3,0)  [var,label=left:]{};
        \node (vl3) at (1.2,-.2)  [var,label=left:]{};
        
        \node (vu1) at (1.3,2.6)  [var,label=left:]{};
        \node (vu2) at (1.3,2.4)  [var,label=left:]{};
        \node (vu3) at (1.2,2.2)  [var,label=left:]{};
        
        \node (vI1) at (-2,3)  [var,label=left:]{};
        \node (vI2) at (-2,2.6)  [var,label=left:]{};
     
        \draw [-] (-2.5,3) to (vI1);
        \draw [-] (vI1) to (vO1);
        \draw [-] (vI1) to (vI2) ;
        \draw [->] (vI2) to (vu1_in);
        \draw [->] (vI2) to (pi_in);

        \path[draw,->] 
        (pi_out.south) -- ++(0,-.23) -- (vl1_in);

        \path[draw,->] 
        (pil_in.west) -- ++(-.2,0) -- ++(0,-1.8) -- (vl2_in);

        \path[draw,->] 
        (piu_in.west) -- ++(-.4,0) -- ++(0,1.4) -- (vu2_in);

        \draw [-] (vu1_out) to (vu1);         
        \draw [-] (vu2_out) to (vu2);
        \draw [-] (vu3_out) to (vu3);
        
        \draw [-] (vl1_out) to (vl1);         
        \draw [-] (vl2_out) to (vl2);
        \draw [-] (vl3_out) to (vl3);    
        
        \draw [-] (vl3) to (vO3);            
        \draw [-] (vu3) to (vO2);

        \path[draw,->] 
        (vu3.south) -- ++(0,-.55) -- (dl_out);

        \path[draw,->] 
        (vl3.north) -- ++(0,.95) -- (du_out);

        \draw [->] (du_in) to (piu_out);   
        \draw [->] (dl_in) to (pil_out);

         \node[draw=black!0,text width=4mm,] at (-.4,-1) {(a)};
        
        
        
        \node (v1) at (0+3,0+.4) [var1,label=left:$$] {};
        \node (v2) at (1.2+3,0+.4) [var1,label=right:$$] {};
        \node (v3) at (2.4+3,0+.4) [var1,label=right:$$] {};
        \node (c1) at ( .5+3,1.7+.4) [check1] {$N$};
        \node (c2) at (2+3,1.7+.4) [check1] {$N$};
        
        
        \draw [-] (v1) to (c1);
        \draw [-] (v1) to (c2);
        \draw [-] (v1) to (c2); 
        \draw [-] (v2) to (c1); 
        \draw [-] (v2) to (c2);
        \draw [-] (v3) to (c1);
        \draw [-] (v3) to (c2);
        
        \node[draw=black!0,text width=4mm,] at (4.2,-1) {(b)};

        \end{tikzpicture}

		\caption{Classical BCCs as (2,3)-regular ensemble: (a) BCC encoder (b) compact graph.}
	\label{fig:bccencgraph}
\end{figure}

\subsection{An Ensemble of $(d_v,d_c)$-regular GLDPC Codes with Convolutional Code Constraints}

Braided convolutional codes can be viewed as a class of turbo-like codes with parity-feedback between the component encoders, as illustrated in Fig.~\ref{fig:bccencgraph}(a). Since the parity symbols enter the other encoder after a delay of one block of $N$ symbols, BCCs are inherently spatially coupled. An uncoupled version of BCCs can be obtained by removing this delay. Fig.~\ref{fig:bccencgraph}(b) shows a compact graph representation of the uncoupled BCCs, in which the variable nodes represent the different blocks of code symbols and the constraint nodes represent the length $N$ component encoder trellises of rate $2/3$. The permutations occur along the edges of the compact graph. 

Observe that the compact graph of the original BCCs is a fully connected $(2,3)$-regular graph, analogous to the protograph of a $(2,3)$-regular LDPC code. In order to generalize BCCs to larger variable node degrees, we can increase the number of component encoders as well as their number of inputs, resulting in $(3,4)$-regular graphs with overall rate $R=1/4$, $(4,5)$-regular graphs with rate $R=1/5$, and so on. These ensembles, however, are hard to compare due to their different rates. Alternatively, we can add edges to the $(2,3)$-regular graph, obtaining $(4,6)$ or $(6,9)$-regular graphs without changing the original rate $R=1/3$. In general, using $d_v$ component encoders of rate $(d_c-1)/d_c$ we can construct arbitrary $(d_v,d_c)$-regular CC-GLDPC codes. Moreover, these codes can be directly compared to the corresponding $(d_v,d_c)$-regular LDPC codes, which have the same overall design rate $R=1-d_v/d_c$ and the same length if their lifting factor is set equal to the number of sections in the trellises.

\subsection{Punctured Trellises for Degree $d_c$ Constraint Nodes}
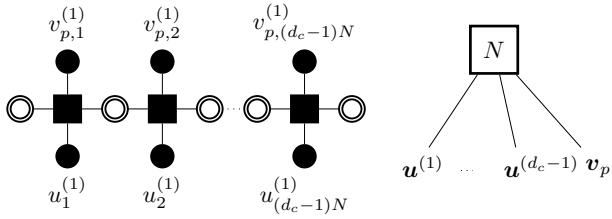
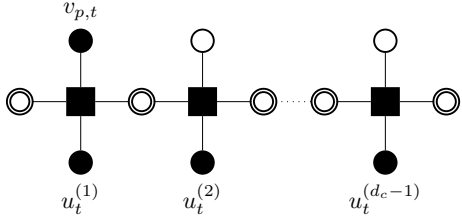
\begin{figure}[t]
	\centering
	\subfigure[Rate $1/2$ trellis (left) in a constraint node (right).]
	{ \scalebox{0.9}{
	        \begin{tikzpicture}[scale=.7]
            \tikzstyle{varnode}=[circle,thick,draw=black!100,fill=black!100,minimum size=1mm]
            \tikzstyle{checknode}=[rectangle,thick,draw=black!100,fill=black!100,minimum size=4mm]
            \tikzstyle{statenode}=[circle,thick,draw=black,double,fill=black!0,minimum size=1mm]
            \tikzstyle{constnode}=[rectangle,line width=.4mm,draw=black!100,fill=black!0,minimum size=7mm]
            \tikzstyle{varcsnode}=[rectangle,thick,draw=black!0,fill=black!0,minimum size=.001]
            
            \node (vu1) at (1,1) [varnode,label=above:$v_{p,1}^{(1)}$] {};
            \node (c1) at  (1,0) [checknode] {};
            \node (vl1) at (1,-1) [varnode,label=below:$u_1^{(1)}$] {};
            \node (vs1) at (0,0) [statenode] {};
           
            \draw [-] (vu1) to (c1);
            \draw [-] (vl1) to (c1);
            \draw [-] (vs1) to (c1); 
    
            \node (vu2) at (2+1,1) [varnode,label=above:$v_{p,2}^{(1)}$] {};
            \node (c2) at  (2+1,0) [checknode] {};
            \node (vl2) at (2+1,-1) [varnode,label=below:$u_2^{(1)}$] {};
            \node (vs2) at (2+0,0) [statenode] {};
           
            \draw [-] (vs2) to (c1);
            \draw [-] (vu2) to (c2);
            \draw [-] (vl2) to (c2);
            \draw [-] (vs2) to (c2); 
            \node (vs3) at (2+2,0) [statenode] {};
            \draw [-] (vs3) to (c2);

            \node (vs4) at (5,0) [statenode] {};
            \draw [dotted] (vs3) -- (vs4);
            \node (vu4) at (6,1) [varnode,label=above:$v_{p,(d_c-1)N}^{(1)}$] {};
            \node (c4) at  (6,0) [checknode] {};
            \node (vl4) at (6,-1) [varnode,label=below:$u_{(d_c-1)N}^{(1)}$] {};
            \draw [-] (vs4) to (c4);
            \draw [-] (vu4) to (c4);
            \draw [-] (vl4) to (c4);
            \node (vs5) at (7,0) [statenode] {};
            \draw [-] (c4) to (vs5);

            \node (cs1) at (10,1.25) [constnode,label=above:$$] {$N$};
            \node (vcs1) at (8.5,-1) [varcsnode,label=above:$$] {};
            \node (vcs2) at (10.5,-1) [varcsnode,label=above:$$] {};
            \node (vcs3) at (12,-1) [varcsnode,label=above:$$] {};
            
            \node (vcs11) at (8.5,-1.8)  [varcsnode,label=above:$\boldsymbol{u}^{(1)}$] {};
            \node (vcs21) at (11,-1.8) [varcsnode,label=above:$\boldsymbol{u}^{(d_c-1)}$] {};
            \node (vcs31) at (12.2,-1.8) [varcsnode,label=above:$\boldsymbol{v}_{p}$] {};
            
            \draw [-] (vcs1) to (cs1);
            \draw [-] (vcs2) to (cs1);
            \draw [-] (vcs3) to (cs1);
            
            \draw [dotted] (9.3,-1.3) to (9.55,-1.3);            
            
            \end{tikzpicture}
    		\label{fig:r12trellis}
    	}}
	\hfill
	\subfigure[Trellis section at time $t$ after puncturing to rate $(d_c-1)/d_c$.]
	{\scalebox{0.9}{
    \begin{tikzpicture}[scale=.9]
    \tikzstyle{varnode}=[circle,thick,draw=black,fill=black!100,minimum size=1.5mm]
    \tikzstyle{pvarnode}=[circle,thick,draw=black,fill=black!0,minimum size=1.5mm]
    \tikzstyle{checknode}=[rectangle,thick,draw=black!100,fill=black!100,minimum size=4mm]
    \tikzstyle{statenode}=[circle,thick,draw=black,double,fill=black!0,minimum size=3mm]
            
            \node (vu1) at (1,1) [varnode,label=above:${v}_{p,t}$] {};
            \node (c1) at  (1,0) [checknode] {};
            \node (vl1) at (1,-1) [varnode,label=below:${u}_t^{(1)}$] {};
            \node (vs1) at (0,0) [statenode] {};
           
            \draw [-] (vu1) to (c1);
            \draw [-] (vl1) to (c1);
            \draw [-] (vs1) to (c1); 
    
            \node (vu2) at (2+1,1) [pvarnode,label=above:$$] {};
            \node (c2) at  (2+1,0) [checknode] {};
            \node (vl2) at (2+1,-1) [varnode,label=below:${u}_t^{(2)}$] {};
            \node (vs2) at (2+0,0) [statenode] {};
           
            \draw [-] (vs2) to (c1);
            \draw [-] (vu2) to (c2);
            \draw [-] (vl2) to (c2);
            \draw [-] (vs2) to (c2); 
            \node (vs3) at (2+2,0) [statenode] {};
            \draw [-] (vs3) to (c2); 
         
            \node (vs4) at (5,0) [statenode] {};
            \draw [dotted] (vs3) to (vs4);
            
            \node (vu3) at (2+1+1+2,1) [pvarnode,label=above:$$] {};
            \node (c3) at  (2+1+1+2,0) [checknode] {};
            \node (vl3) at (2+1+1+2,-1) [varnode,label=below:${u}_t^{(d_c-1)}$] {};
            \node (vs5) at (5+2,0) [statenode] {};
            
            \draw [-] (vs5) to (c3);
            \draw [-] (vs4) to (c3);
            \draw [-] (vu3) to (c3);
            \draw [-] (vl3) to (c3);
            
    \end{tikzpicture}

	}}
	\\
		\caption{Factor graph representation of a constraint node trellis.}\label{fig:puncturedtrellis}
	\label{fig:constnode}
\end{figure}

Consider a degree $d_c$ constraint node in the compact graph. Each edge represents a length $N$ sequence of code symbols that are represented by the connected variable node. One of these sequences will correspond to the parity sequence $\boldsymbol{v}_p=(v_{p,1},\dots,v_{p,N})$ of the component convolutional code and the other $d_c-1$  to the information sequences $\boldsymbol{u}^{(i)}=(u_{1}^{(i)},\dots,u_{N}^{(i)})$, $i=1,\dots,d_c-1$.

In order to construct a component code of rate $(d_c-1)/d_c$, for any $d_c \geq 2$ we can use a rate-$1/2$ mother code with a trellis of $(d_c-1)N$ sections. A factor graph of such a trellis is shown in Fig.~\ref{fig:puncturedtrellis}. The desired code rate is achieved by puncturing $d_c-2$ of the parity bits in each segment of $d_c-1$ trellis sections, as shown by white circles in the factor graph. For example, the degree $d_c=3$ constraint node of the classical BCC graph in Fig.~\ref{fig:bccencgraph}(b) can be implemented by a trellis of length $2N$ in which every second parity bit is punctured to achieve a rate-$2/3$ component encoder. In general, the puncturing patterns in different segments of the trellis can be time-varying, and finding patterns that optimize the thresholds or the distance spectrum of the resulting codes is an open problem. In our threshold analysis, we will assume uniform random puncturing within each segment, such that a parity bit remains unpunctured with probability $1/(d_c-1)$.     

The strength of the constraint nodes can be flexibly changed without altering their degree by simply increasing the number of states in the component code trellis. In this work, we consider recursive systematic convolutional encoders with generator polynomials $(1,1/3)$, $(1,5/7)$ and $(1,13/15)$ in octal notation, with 2, 4 and 8 trellis states, respectively.


\subsection{Single-Edge Type Ensembles}

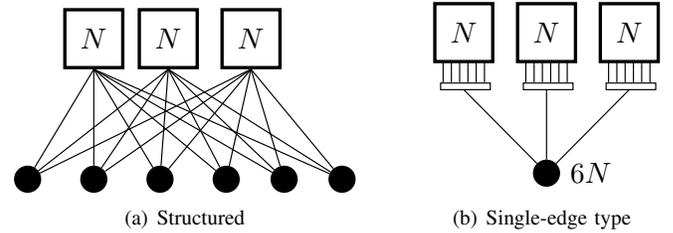
\begin{figure}[th]
	\centering
	\subfigure[Structured]
	{\scalebox{1.1}{
	\begin{tikzpicture}
        [var/.style={circle,draw=black!100,fill=black!100,thick,
        inner sep=0pt,minimum size=3mm},
        check/.style={rectangle,draw=black!100,fill=black!0,line width=.4mm,
        inner sep=0pt,minimum size=7mm}]
        
        
        \node (v1) at (.7+0,0) [var] {};
        \node (v2) at (.8+.7,0) [var] {};
        \node (v3) at (.9+1.4,0) [var] {};
        \node (v4) at (1+2.1,0) [var] {};
        \node (v5) at (1+2.8,0) [var] {};
        \node (v6) at (1+3.5,0) [var] {};
        
        \node (c1) at (1+.2+.3,1.7) [check] {$N$};
        \node (c2) at (1+1.1+.3,1.7) [check] {$N$};
        \node (c3) at (1+2.1+.3,1.7) [check] {$N$};

        
        \draw [-] (v1) to (c1.south);
        \draw [-] (v1) to (c2.south);
        \draw [-] (v1) to (c3.south); 

        \draw [-] (v2) to (c1.south);
        \draw [-] (v2) to (c2.south);
        \draw [-] (v2) to (c3.south); 

        \draw [-] (v3) to (c1.south);
        \draw [-] (v3) to (c2.south);
        \draw [-] (v3) to (c3.south); 

        \draw [-] (v4) to (c1.south);
        \draw [-] (v4) to (c2.south);
        \draw [-] (v4) to (c3.south); 

        \draw [-] (v5) to (c1.south);
        \draw [-] (v5) to (c2.south);
        \draw [-] (v5) to (c3.south); 

        \draw [-] (v6) to (c1.south);
        \draw [-] (v6) to (c2.south);
        \draw [-] (v6) to (c3.south); 

        \end{tikzpicture}

		\label{fig:36graph}
	}}
\hfill
\subfigure[Single-edge type]
{ \scalebox{1.1}{
	\begin{tikzpicture}
        [var/.style={circle,draw=black!100,fill=black!100,thick,inner sep=0pt,minimum size=3mm},
        check/.style={rectangle,draw=black!100,fill=black!0,line width=.4mm,inner sep=0pt,minimum size=7mm}]
        \tikzstyle{invnode}=[circle,thick,draw=black!100,fill=black!0,minimum size=1.5mm]
        \def\permuter{-- +(.15,-.1)}
        \def\splitterpath{-- +(.6,0) -- +(.6,.08) -- +(0,.08) -- cycle}
        
        
        \node (v1) at (-2,0) [var,label=right:$6N$] {};
       
        \node (c1) at ( -3,1.7) [check] {$N$};
        \node (c2) at (-2,1.7) [check] {$N$};
        \node (c3) at (-1,1.7) [check] {$N$};

        
        \draw [-] (v1) to (-3-0,1.01) ;
        \draw [-] (v1) to (-2-0,1.01);
        \draw [-] (v1) to (-1-0,1.01); 

        \draw (-3-.28,1.01) [fill=white] \splitterpath {};
        \draw (-2-.28,1.01) [fill=white] \splitterpath {};
        \draw (-1-.28,1.01) [fill=white] \splitterpath {};
        
        \foreach \x in {0,1,2,3,4,5}
            {
                \draw [-] (-3.25+\x*.1,1.1) to (-3.25+\x*.1,1.35);
                \draw [-] (-2.25+\x*.1,1.1) to (-2.25+\x*.1,1.35);
                \draw [-] (-1.25+\x*.1,1.1) to (-1.25+\x*.1,1.35);
            }
        

        \end{tikzpicture}

	\label{fig:46_sim}
}}
\caption{Graph represenations of (3,6)-regular ensembles.}
\label{fig:modelsim}
\end{figure}

The compact graph of a $(3,6)$-regular ensemble is shown in Fig.~\ref{fig:modelsim}(a). Since the edges define a clear assignment from outputs of the constraint nodes to the variable nodes, this is an example of a structured graph. In order to simplify analysis, when computing ensemble weight enumerators and thresholds, we will instead consider single-edge type ensembles like shown in Fig.~\ref{fig:modelsim}(b).

\section{Finite-Length Ensemble Weight Enumerators}
A weight enumerator analysis for different ensembles of turbo-like codes, including uncoupled BCCs, was carried out in \cite{MOSCTCNewtrade}. Considering the $(2,3)$-regular ensemble in Fig.~\ref{fig:bccencgraph}(b), let $A^{(j)}_{i_1,i_1,p}$ denote the number of code sequences of input weights $i_1$, $i_2$ and parity weight $p$ for the length $N$ convolutional code at constraint node $j$. Assuming uniform random permutations, it is then possible to compute the average number of codewords $\bar{A}^{BCC}_{i,p}$ over all codes in the ensemble as follows:
\begin{equation} \label{eq:avWEstruct}
    \bar{A}_{i,p}^{BCC} = \sum_{p_1} \frac{A^{(1)}_{i,p_1,p-p_1} \cdot A^{(2)}_{i,p-p_1,p_1}}{\binom{N}{i} \binom{N}{p_1} \binom{N}{p-p_1} } \ . 
\end{equation}
In principle, a generalization to general structured $(d_v,d_c)$ ensembles, like illustrated in Fig.~\ref{fig:modelsim}(a), is possible\footnote{As pointed out in \cite{MOSCTCNewtrade}, the weight enumerator expression in \eqref{eq:avWEstruct} is equivalent to protograph-based GLDPC code ensembles analyzed in \cite{AbuSurra2011}.}. Unfortunately, this approach becomes numerically infeasible for a given $N$ when the variable node degree increases. In this work, we consider unstructured single-edge type ensembles, like illustrated in Fig.~\ref{fig:modelsim}(b), and generalize Gallager's weight enumerator analysis for LDPC codes \cite{GallagerLDPCBook,BocharovaSpectraLDPC} to our ensembles:
\begin{equation} \label{eq:avWE}
    \bar{A}_w^{(d_v,d_c)} =  \frac{\left(A_w^{(j)}\right)^{d_v}}{{\binom{d_c \cdot N}{w}}^{d_v-1}} \ ,
\end{equation}
where $w$ is the total weight of information and parity bits.



A bound on the minimum distance $d_{min}$ of codes in an ensemble can be obtained by computing the largest positive integer $\hat{d}$ that satisfies the expression
\begin{equation} \label{eq:mindb}
    \sum_{w=1}^{\hat{d}-1}  \bar{A}_w^{(d_v,d_c)} < 1-\alpha 
\end{equation}
for a given $\alpha<1$. Then a fraction $\alpha$ of all codes in the ensemble must have a minimum distance $d_{min} \geq \hat{d}$.


\section{Convergence Thresholds for the BEC}
We assume that the BP decoder of a CC-GLDPC code is based on optimal bitwise a-posteriori probability (APP) decoding at the constraint nodes\footnote{This is the equivalent to the classical turbo decoder \cite{McEliece1998}.}. The MAP decoding threshold, on the other hand, refers to the optimal bitwise decoding of the overall code, which is computationally infeasible. For the BEC it is possible to find analytical expressions for the input/output transfer functions of the component decoders \cite{ASTEExitModelP}. By means of these it is possible to derive exact DE equations for the ensembles introduced in Section~II, which capture the evolution of erasure probabilities of  messages being passed back and forth along the edges in the graph.

\subsection{Density Evolution Equations for Uncoupled Ensembles}
Consider a $(d_v,d_c)$-regular graph and let $e_{j,k}$ denote the edge connecting variable node $j$ to constraint node $k$, where $j \in \{1,\dots,d_v\}$ and $k \in \{1,\dots,d_c\}$. 
The DE update at a constraint node in iteration $i$ can be expressed as
\begin{equation}
p^{(i)}(e_{j,k}) = f_k\left(q^{(i-1)}(e_{j,1}),\dots,q^{(i-1)}(e_{j,d_c})\right) \ , 
\end{equation}
where $q^{(i)}(e_{j,k})$ and $p^{(i)}(e_{j,k})$ denote the probabilities that messages passed from variable to check nodes and from check nodes to variable nodes are erased, respectively. $f_k$ denotes the (extrinsic) transfer function of the constraint node, corresponding to the trellis output message type associated with edge $e_{j,k}$. For conventional LDPC codes this transfer function is independent of $k$ and reduces to the well-known expression
\begin{align}
p^{(i)}(e_{j,k}) &= 1 - \prod_{k' \setminus k} \left(1 - q^{(i-1)}(e_{j,k'})\right) \\ &= 1-\left(1 - q^{(i-1)} \right)^{d_c-1} \ .
\end{align}
Before the first iteration $i=1$, all input erasure probabilities are initialized to $q^{(0)}(e_{j,k})=\epsilon$, which is the erasure probability of the BEC. At a variable node, the DE update  can be written as
\begin{equation}\label{eq:DEvariable}
q^{(i)}(e_{j,k}) = \epsilon \cdot \prod_{j' \setminus j} p^{(i)}(e_{j',k}) = \epsilon \cdot \left( p^{(i)} \right)^{d_v-1} . 
\end{equation}
In this work we consider single-edge type regular graphs, as illustrated in Fig.~\ref{fig:modelsim}(b), for which the trellis outputs of the constraint nodes are distributed uniformly over all code symbols of the variable node. In this case $p^{(i)}$ and $q^{(i-1)}$ are equal along all edges of the graph.

\subsection{Transfer Functions for Punctured Trellises}
In order to compute the transfer functions of the constraint nodes of the graph, a rate-1/2 trellis is punctured to match the constraint node degree of the graph. The mother code transfer functions for the considered generator polynomials can be  derived as shown in \cite{Moloudi_SCTC_Journal}. Let $f_\text{s}$ and $f_\text{p}$ denote these transfer functions for systematic and parity bits, respectively. Then we can write
\begin{align}
    p_\text{s}^{(i)} &= f_\text{s}\left(q_\text{s}^{(i)},q_\text{p}^{(i)}\right) \ , \label{Eq:tfnonpunc1} \\
    p_\text{p}^{(i)} &= f_\text{p}\left(q_\text{s}^{(i)},q_\text{p}^{(i)}\right) \ , \label{Eq:tfnonpunc2}
\end{align}
where $p_\text{s}^{(i)}$ and $p_\text{p}^{(i)}$ denote the extrinsic output erasure probabilities, and $q_\text{s}^{(i)}$ and $q_\text{p}^{(i)}$ the erasure probabilities of incoming messages to the constraint node. Assuming that random puncturing of parity bits is used for achieving a target rate $(d_c-1)/d_c$, these input erasure probabilities are given by
\begin{align}
    q_\text{s}^{(i)} &= q^{(i-1)} \ , \\ 
    q_\text{p}^{(i)} &= \frac{d_c-2}{d_c-1} \cdot 1 + \frac{1}{d_c-1} \cdot q^{(i-1)} = \frac{q^{(i-1)}+d_c-2}{d_c-1} \ . 
\end{align}
The average erasure probability of messages sent from the constraint nodes to the variable node is equal to
\begin{equation}\label{eq:wep}
    p^{(i)}=\frac{(d_c-1)}{d_c} \cdot p_\text{s}^{(i)} + \frac{1}{d_c} \cdot p_\text{p}^{(i)}  \ .
\end{equation}
DE iteration $i$ is then completed by a variable note update according to \eqref{eq:DEvariable}, resulting in $q^{(i)}$.

\subsection{Density Evolution Equations for Coupled Ensembles}

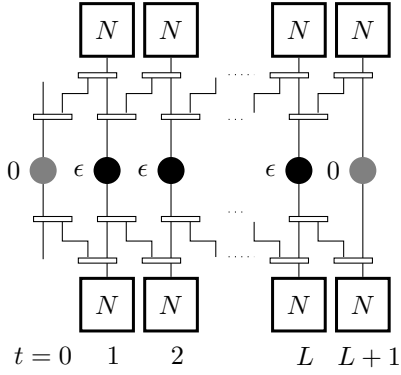
\begin{figure}[t]
	\centering

\begin{tikzpicture}[scale= .85]
\tikzstyle{varnode}=[circle,thick,draw=black,fill=black!100,minimum size=1.5mm]
\tikzstyle{checknode}=[rectangle,line width=.4mm,draw=black!100,fill=black!0,minimum size=7mm]
\tikzstyle{splitter}=[rectangle,thick,draw=white,fill=white!0,minimum size=.005mm]
\def\splitterpath{-- +(.6,0) -- +(.6,.08) -- +(0,.08) -- cycle}
\tikzstyle{invnode}=[circle,thick,draw=black!0,fill=black!0,minimum size=1.5mm]
\tikzstyle{tnode}=[circle,thick,draw=black!50,fill=black!50,minimum size=1.5mm]

\node (t1) at (0,.5) [invnode, label=below:{$t=0$}] {};
\node (vu0) at (0,2.9) [tnode, label=left:$0$] {};
\node (cu01) at (0,4.5) [invnode, label=below:$$] {};
\node (cu02) at (0,1.3) [invnode, label=below:$$] {};
\node (s01) at (.3,2.25) [splitter, label=below:$$] {};
\node (s02) at (.3,3.65) [splitter, label=below:$$] {};

\draw [-] (vu0) to (cu01);
\draw [-] (vu0) to (cu02);

\draw (-.15,3.7) [fill=white] \splitterpath {};
\draw (-.15,2.1) [fill=white] \splitterpath {};

\path[draw,-,black] 
    (s01.south) -- ++(0,-.3) -- ++(.4,0) -- ++(0,-.25) ;
\path[draw,-,black] 
    (s02.north) -- ++(0,.3) -- ++(.4,0) -- ++(0,.25) ;

\node (t1) at (1+.1,.5) [invnode, label=below:{$1$}] {};
\node (vu1) at (1,5.1) [checknode, label=above:$$] {$N$};
\node (s11) at (1.3,2.25) [splitter, label=below:$$] {};
\node (cu1) at (1,3-.1) [varnode, label=left:$\epsilon$] {$$};
\node (s12) at (.85+.45,3.65) [splitter, label=below:$$] {};
\node (vl1) at (1,3-2.2) [checknode, label=below:$$] {$N$};

\draw [-] (vu1) to (cu1);
\draw [-] (vl1) to (cu1);

\draw (.55,4.35) [fill=white] \splitterpath {};
\draw (.85,3.7) [fill=white] \splitterpath {};
\draw (.85,3-.9) [fill=white] \splitterpath {};
\draw (.55,3-1.55) [fill=white] \splitterpath {};

\path[draw,-,black] 
    (s11.south) -- ++(0,-.3) -- ++(.4,0) -- ++(0,-.25) ;
\path[draw,-,black] 
    (s12.north) -- ++(0,.3) -- ++(.4,0) -- ++(0,.25) ;

\node (t2) at (1+1+.1,.5) [invnode, label=below:$2$] {};
\node (vu2) at (1+1,5.1) [checknode, label=above:$$] {$N$};
\node (s21) at (1.3+1,2.25) [splitter, label=below:$$] {};
\node (cu2) at (1+1,3-.1) [varnode, label=left:$\epsilon$] {$$};
\node (s22) at (.85+.45+1,3.65) [splitter, label=below:$$] {};
\node (vl2) at (1+1,3-2.2) [checknode, label=below:$$] {$N$};

\draw [-] (vu2) to (cu2);
\draw [-] (vl2) to (cu2);

\draw (.55+1,4.35) [fill=white] \splitterpath {};
\draw (.85+1,3.7) [fill=white] \splitterpath {};
\draw (.85+1,3-.9) [fill=white] \splitterpath {};
\draw (.55+1,3-1.55) [fill=white] \splitterpath {};

\path[draw,-,black] 
    (s21.south) -- ++(0,-.3) -- ++(.4,0) -- ++(0,-.25) ;
\path[draw,-,black] 
    (s22.north) -- ++(0,.3) -- ++(.4,0) -- ++(0,.25) ;


\draw [dotted] (2.9, 3.7) to (3.3,3.7);
\draw [dotted] (2.9, 2.25) to (3.3,2.25);

\draw [dotted] (2.9, 3.9+.5) to (3.3,3.9+.5);
\draw [dotted] (2.9, 2.05-.5) to (3.3,2.05-.5);


\node (sd1) at (1.3+2,2.25) [splitter, label=below:$$] {};
\node (sd2) at (.85+.45+2,3.65) [splitter, label=below:$$] {};
\path[draw,-,black] 
    (sd1.south) -- ++(0,-.3) -- ++(.4,0) -- ++(0,-.25) ;
\path[draw,-,black] 
    (sd2.north) -- ++(0,.3) -- ++(.4,0) -- ++(0,.25) ;

\node (tL) at (1+3+.1,.5) [invnode, label=below:$L$] {};
\node (vu3) at (1+3,5.1) [checknode, label=above:$$] {$N$};
\node (s31) at (1.3+3,2.25) [splitter, label=below:$$] {};
\node (cu3) at (1+3,3-.1) [varnode, label=left:$\epsilon$] {$$};
\node (s32) at (.85+.45+3,3.65) [splitter, label=below:$$] {};
\node (vl3) at (1+3,3-2.2) [checknode, label=below:$$] {$N$};

\draw [-] (vu3) to (cu3);
\draw [-] (vl3) to (cu3);

\draw (.55+3,4.35) [fill=white] \splitterpath {};
\draw (.85+3,3.7) [fill=white] \splitterpath {};
\draw (.85+3,3-.9) [fill=white] \splitterpath {};
\draw (.55+3,3-1.55) [fill=white] \splitterpath {};

\path[draw,-,black] 
    (s31.south) -- ++(0,-.3) -- ++(.4,0) -- ++(0,-.25) ;
\path[draw,-,black] 
    (s32.north) -- ++(0,.3) -- ++(.4,0) -- ++(0,.25) ;

\node (tLp1) at (1+3+1+.1,.5) [invnode, label=below:$L+1$] {};
\node (vu4) at (1+3+1,5.1) [checknode, label=right:$$] {$N$};
\node (s41) at (1.3+3+1,4.5) [splitter, label=below:$$] {};
\node (cu4) at (1+3+1,3-.1) [tnode, label=left:$0$] {$$};
\node (s42) at (.85+.45+3+1,1.4) [splitter, label=below:$$] {};
\node (vl4) at (1+3+1,3-2.2) [checknode, label=right:$$] {$N$};

\draw [-] (vu4) to (cu4);
\draw [-] (vl4) to (cu4);

\draw (.55+3+1,4.35) [fill=white] \splitterpath {};
\draw (.55+3+1,3-1.55) [fill=white] \splitterpath {};
\end{tikzpicture}

\caption{Spatially coupled $(2,3)$-regular ensemble: single-edge type representation with coupling memory $m=1$. }\label{fig:coupledgraph}
\end{figure}

For spatially coupled ensembles, as illustrated in Fig.~\ref{fig:coupledgraph}, we have a sequence of $L$ graphs whose constraint nodes and variable nodes are placed at time instants $t=1,\dots,L$. We consider ensembles with uniform coupling, i.e., every edge from a variable node at time $t$ is connected to a constraint node at time $t'\in\{t,t+1,\dots,t+m\}$ with probability $1/(m+1)$, where $m$ is called the coupling memory.

For conventional SC-LDPC codes a constraint node represents a rate $(d_c-1)/d_c$ single parity-check code and the update equation becomes
\begin{equation}
p_t^{(i)} = 1-\left(1 - \frac{1}{m+1} \sum_{\ell=0}^m q_{t-\ell}^{(i-1)} \right)^{d_c-1} \ .
\end{equation}
For CC-GLDPC codes with punctured component code trellises, we update the transfer functions \eqref{Eq:tfnonpunc1}--\eqref{Eq:tfnonpunc2} with the input erasure probabilities
\begin{align}
    q_{\text{s},t}^{(i)} &= \frac{1}{m+1} \sum_{\ell=0}^m q_{t-\ell}^{(i-1)} \ , \\ 
    q_{\text{p},t}^{(i)} &= \frac{q_{\text{s},t}^{(i)}+d_c-2}{d_c-1} \  
\end{align}
and obtain $p_t^{(i)}$ analogously to \eqref{eq:wep} for each $t$. Before the first iteration $i=1$, the input erasure probabilities are initialized to $q_t^{(0)}=\epsilon$ for $t \in \{1,\dots,L\}$. For all other $t$, the code symbols are known to be zero by definition and  $q_t^{(0)}=0$.

At the variable nodes, the DE update can be written as
\begin{equation}
    q_t^{(i)} = \epsilon \cdot \left( \frac{1}{m+1} \sum_{\ell=0}^m p_{t+\ell}^{(i)} \right)^{d_v-1} \ , \quad t=1,\dots,L \ .
\end{equation}

\subsection{BP and MAP Thresholds}
The BP threshold $\epsilon_{BP}$ is defined as the largest channel erasure probability $\epsilon$ for which the a-posteriori erasure probabilities $p_{a}^{(i)}$ at the output of the BP decoder converge to zero for all variable nodes as the number of iterations $i$ tends to infinity. The probabilities  $p_{a}^{(i)}=\epsilon \cdot \left( p^{(i)} \right)^{d_v}$ can be computed by repeated use of the density evolution equations for different $\epsilon$. The (bitwise) MAP threshold $\epsilon_{MAP}$ can obtained by applying the area theorem \cite{ASTEExitModelP,Measson2009}, which makes it possible to connect the performance under BP decoding to that of MAP decoding.
Let $\bar{p}_{e}(\epsilon)=\lim_{i \rightarrow \infty} \bar{p}_{a}(\epsilon)^{(i)}/\epsilon$ denote the average extrinsic probability of erasure. An upper bound on the MAP threshold can be computed by the equation
\begin{equation}
\int_{\epsilon_{MAP}}^{1} \bar{p}_{e}(\epsilon) d\epsilon = R \ , 
\end{equation}
where $R$ is the rate of the considered code.

\section{Results and Discussion}

\begin{figure}[t]
	\centerline{\includegraphics[scale=.8]{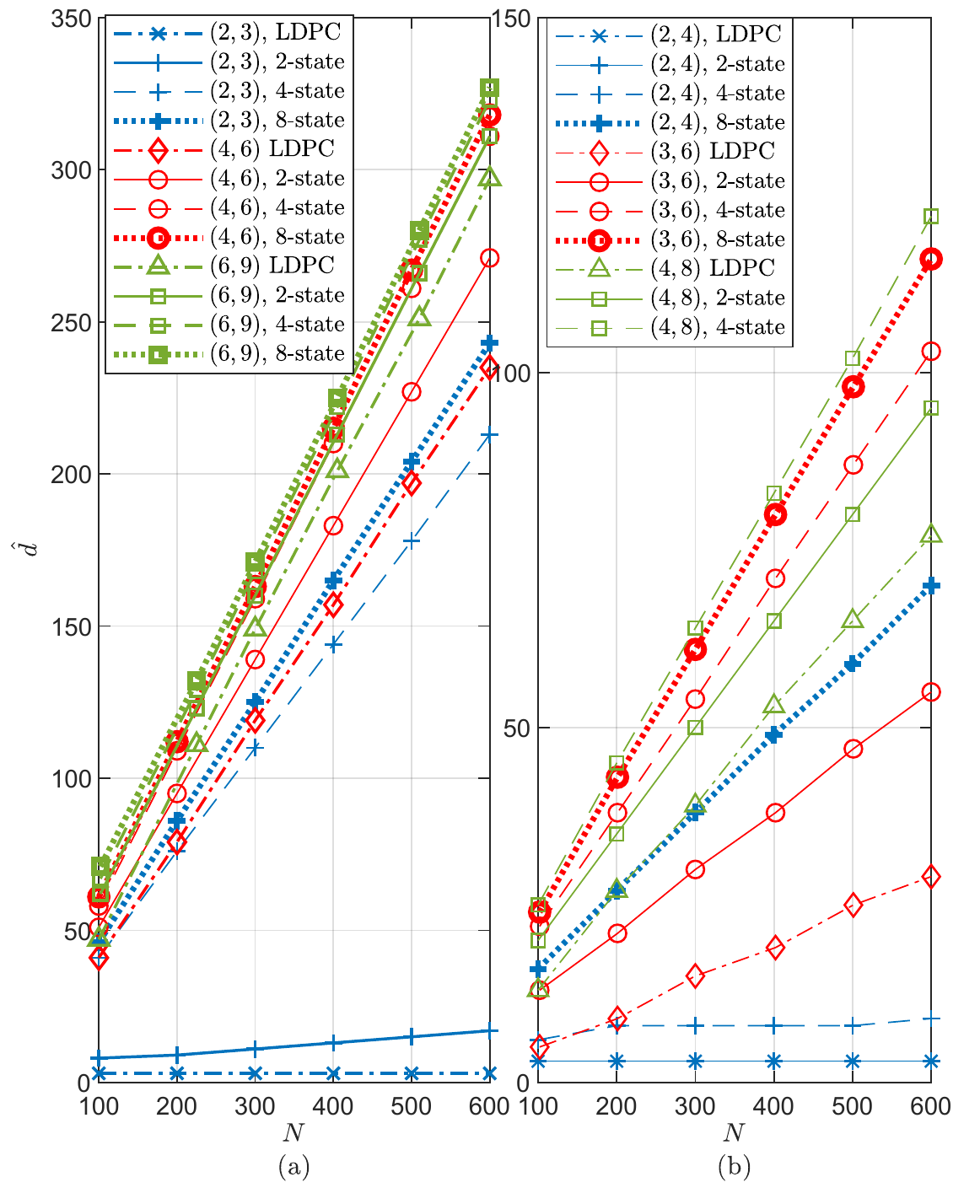}}\vspace*{-3mm}
	\caption{Bound on the minimum distance: a fraction $\alpha=1/2$ of codes in an ensemble have $d_{min} \geq \hat{d} $: (a)  $R=1/3$ (b) $R=1/2$.}
	\label{fig:mindbound}
\end{figure}

\subsection{Minimum Distance Bounds}
The minimum distance bounds, computed using \eqref{eq:mindb} with $\alpha=0.5$ for the ensembles of rate $R=1/3$ and $R=1/2$, are shown in Fig.~\ref{fig:mindbound}(a) and (b), respectively. It follows from the bound that half of the codes in an ensemble must have minimum distance $d_{min} \geq \hat{d}$. 
From the figure, it is observed that in general the minimum distance improves when the component code gets stronger. Furthermore, for a given component code the distance improves if the variable node is increased. Interestingly,  for $R=1/3$, the weakest CC-GLDPC codes with 2-state components appear to have better minimum distance than classical LDPC codes. The results show that we indeed can reduce the variable node degree if we increase the number of states of the component encoder. For example, codes from the $(3,6)$ ensemble with 4 states have better minimum distance than those of the $(4,8)$ ensemble with 2 states and the $(4,8)$ LDPC ensemble. As expected \cite{GallagerLDPCBook}, the minimum distances of LDPC ensembles with variable node degree 2 are very poor, which is also observed for the 2-state CC-GLDPC ensembles.

\subsection{Thresholds}

Table~\ref{Table:ThresUC} shows the BP thresholds and MAP thresholds for the uncoupled ensembles of rate $R=1/3$ and $R=1/2$. It is observed that BP thresholds tend to {\em decrease} with increasing variable node degree and increasing number of trellis states. However, MAP thresholds tend to {\em increase} with increasing variable node degree and increasing number of trellis states. An exception from this behavior is observed for 2-state ensembles and LDPC ensembles with variable node degree 2 at rate $R=1/2$.\footnote{These ensembles are poor and not of practical interest. As shown in \cite{LentmaierBERPIterDec}, regular GLDPC ensembles with $d_v<3$ require component codes with minimum distance $d_{min}>2$ in order to guarantee that the block error probability tends to zero at the BP threshold.}

\begin{table}[t]
	\caption{Thresholds of uncoupled ensembles.}
	\begin{center}
		\begin{tabular}{lcrrr}
            \hline
            \textbf{Thresholds}& \textbf{States} &\multicolumn{3}{c}{\textbf{Graph rate $\boldsymbol{1/3}$}}  \\
                        &        &  $\boldsymbol{(2,3)}$ & $\boldsymbol{(4,6)}$ & $\boldsymbol{(6,9)}$     \\
            \hline
             $\epsilon_{BP}$  &2 & 0.6086  & 0.5339 & 0.4698 \\
             $\epsilon_{MAP}$ &2 & 0.6213 & 0.6564 & 0.6610 \\
            \hline
            $\epsilon_{BP}$ & 4& 0.5618 & 0.4464  & 0.3853 \\
            $\epsilon_{MAP}$ &4& 0.6647  & 0.6662  & 0.6664\\
            \hline
            $\epsilon_{BP}$& 8& 0.5352 & 0.4041 & 0.3401  \\
            $\epsilon_{MAP}$& 8& 0.6659 & 0.6665  & 0.6666      \\
             \hline
            $\epsilon_{BP}$ &LDPC& 0.2570   & 0.5061  & 0.4034  \\
            $\epsilon_{MAP}$ &LDPC& 0.5089  & 0.6658 & 0.6667  \\
            \hline
            \textbf{}& \textbf{} &\multicolumn{3}{c}{\textbf{Graph rate $\boldsymbol{1/2}$}}  \\
                        &        &  $\boldsymbol{(2,4)}$ & $\boldsymbol{(3,6)}$ & $\boldsymbol{(4,8)}$     \\
             \hline
            $\epsilon_{BP}$ &2& 0.3234              &   0.4110             &    0.3916   \\
            $\epsilon_{MAP}$ &2  &   0.3444            &    0.4557        &   0.4737\\
            \hline
            $\epsilon_{BP}$ & 4& 0.4426 & 0.3929 & 0.3555  \\
            $\epsilon_{MAP}$ &4 & 0.4890 & 0.4958  & 0.4976 \\
            \hline
            $\epsilon_{BP}$ & 8 &   0.4249      &  0.3638     &   0.3225  \\
            $\epsilon_{MAP}$ &8 &   0.4955      &  0.4985     &  0.4991       \\
             \hline
            $\epsilon_{BP}$ &LDPC& 0.1725   & 0.4294  &  0.3834  \\
            $\epsilon_{MAP}$ &LDPC& 0.4002   & 0.4883  & 0.4978  \\
            \hline

        \end{tabular}
		\label{Table:ThresUC}
	\end{center}
\end{table}

In order to understand the trade-off between variable node degree and number of trellis states, we compare the minimum distances, BP thresholds and MAP thresholds of the $(2,3)$ ensemble with 4 states and 8 states, the $(4,6)$ ensemble with 4 states, and the $(6,9)$ ensemble with 2 states. The $(2,3)$ ensemble with 8 states has better BP and MAP thresholds than the $(6,9)$ ensemble with 2 states, but the minimum distance is clearly worse. If a comparable minimum distance is a requirement, but it is desired to keep the variable node degree as low as possible, then the $(4,6)$ ensemble with 4 states can be used instead. In terms of the BP decoding performance, this ensemble is not as good as the $(2,3)$ ensemble with 8 states or the $(6,9)$ ensemble with 2 states, but it has the best MAP threshold among these three ensembles. The strong MAP threshold of the $(4,6)$ ensemble with 4 states makes it a compelling candidate for spatial coupling.

The BP thresholds of the spatially coupled ensembles are shown in Table~\ref{Table:thdsc} for different coupling memories $m$, until saturation to the MAP threshold occurs. Due to the threshold saturation phenomenon, the BP thresholds approach the MAP thresholds as $m$ increases and the $(4,6)$ ensemble with 4 states from our example above has now a better BP threshold than the other considered ensembles.


\begin{table}[t]
\caption{Thresholds of SC ensembles.}
\begin{center}
\begin{tabular}{lcccccc}
\hline
States/SC & \multicolumn{6}{c}{Graph}                           \\
memory    & $\boldsymbol{(2,3)}$     & $\boldsymbol{(4,6)}$     & $\boldsymbol{(6,9)}$     & $\boldsymbol{(2,4)}$     & $\boldsymbol{(3,6)}$     & $\boldsymbol{(4,8)}$     \\
\hline
2/1     & 0.6212 & 0.6532 & 0.6294 & 0.3345 & 0.4556 & 0.4715 \\
2/2     & -      & 0.6563 & 0.6586 & -      & 0.4557 & 0.4736 \\
2/3     & -      & 0.6563 & 0.6608 & -      & -      & -      \\
2/4     & -      & 0.6564 & 0.6609 & -      & -      & -      \\
\hline
4/1     & 0.6581 & 0.6351 & 0.5822 & 0.4885 & 0.4911 & 0.4829 \\
4/2     & 0.6645 & 0.6639 & 0.6510 & 0.4890 & 0.4956 & 0.4967 \\
4/3     & 0.6647 & 0.6661 & 0.6643 & -      & 0.4957 & 0.4975 \\
4/4     & -      & 0.6662 & 0.6662 & -      & -      & -      \\
4/5     & -      & -      & 0.6663 & -      & -      & -      \\
\hline
8/1     & 0.6479 & 0.6029 & 0.5364 & 0.4917 & 0.4825 & 0.4644 \\
8/2     & 0.6643 & 0.6560 & 0.6271 & 0.4953 & 0.4974 & 0.4947 \\
8/3     & 0.6658 & 0.6651 & 0.6570 & 0.4954 & 0.4983 & 0.4987 \\
8/4     & 0.6659 & 0.6664 & 0.6645 & -      & 0.4984 & 0.4991 \\
8/5     & -      & -      & 0.6662 & -      & -      & -      \\
8/6     & -      & -      & 0.6664 & -      & -      & -      \\
8/7     & -      & -      & 0.6665 & -      & -      & -      \\
\hline
LDPC/1  & 0.5014 & 0.6611 & 0.6118 & 0.3348 & 0.4880 & 0.4943 \\
LDPC/2  & -      & 0.6655 & 0.6622 & -      & 0.4881 & 0.4977 \\
LDPC/3  & -      & 0.6655 & 0.6664 & -      & -      & -      \\
LDPC/4  & -      & 0.6655 & 0.6665 & -      & -      & -      \\
LDPC/5  & -      & 0.6656 & -      & -      & -      & -     \\
\hline
\end{tabular}
\label{Table:thdsc}
\end{center}
\end{table}

\section{Conclusion}
We have introduced a family of GLDPC codes with convolutional code constraints, which allows a one-by-one comparison with corresponding LDPC code ensembles of arbitrary variable node and check node degrees. Although we have focused in this work on regular graphs only, the ensembles can easily be extended to irregular codes by removing some edges in the graphs. Furthermore, it is possible to use component codes of lower rate at the constraint nodes, but then the rate of the resulting ensembles will be different from the LDPC code ensembles defined by the same graphs. An advantage of using convolutional codes at the constraint nodes is that the strength of the component codes can be altered without changing the node degrees in the graph.   

The considered ensembles permit us to study the trade-off between variable node degree and component code strength in terms of their minimum distance, BP decoding thresholds and MAP decoding thresholds.  A larger number of trellis states  is shown to yield better minimum distances and MAP thresholds but degraded BP thresholds. This degraded BP performance is avoided  by applying spatial coupling to the underlying uncoupled ensembles. It can also be seen from the threshold results that for a regular LDPC ensemble with $d_v>2$, there is an alternative CC-GLDPC ensemble, having a lower variable node degree than the LDPC ensemble, that has a better BP threshold and almost similar or better MAP threshold than the LDPC ensemble.


\bibliographystyle{IEEEbib}
\bibliography{IEEEabrv,referencesJ}

\end{document}